# Domain alignment within ferroelectric/dielectric PbTiO$_3$/SrTiO$_3$ superlattice nanostructures†

J. Park,[a] J. Mangeri,[b] Q. Zhang,[a] M. H. Yusuf,[c] A. Pateras,[a] M. Dawber,[c] M. V. Holt,[d] O. G. Heinonen,[e, f] S. Nakhmanson,[b, g] and P. G. Evans[a]

The ferroelectric domain pattern within lithographically defined PbTiO$_3$/SrTiO$_3$ ferroelectric/dielectric heteroepitaxial superlattice nanostructures is strongly influenced by the edges of the structures. Synchrotron x-ray nanobeam diffraction reveals that the spontaneously formed 180° ferroelectric stripe domains exhibited by such superlattices adopt a configuration in rectangular nanostructures in which domain walls are aligned with long patterned edges. The angular distribution of x-ray diffuse scattering intensity from nanodomains indicates that domains are aligned within an angular range of approximately 20° with respect to the edges. Computational studies based on a time-dependent Landau-Ginzburg-Devonshire model show that the preferred direction of the alignment results from lowering of the bulk and electrostrictive contributions to the free energy of the system due to the release of the lateral mechanical constraint. This unexpected alignment appears to be intrinsic and not a result of distortions or defects caused by the patterning process. Our work demonstrates how nanostructuring and patterning of heteroepitaxial superlattices allow for pathways to create and control ferroelectric structures that may appear counterintuitive.

## Introduction

Ferroelectric materials display a remarkable range of phenomena that can be influenced by the formation of nanostructures or nanoscale interfaces.[1] Fundamental size effects are prominent in nanoscale ferroelectric crystals or crystalline epitaxial islands,[1] composites incorporating nanoscale ferroelectric crystallites,[2-3] and in ferroelectric layers with nanoscale thickness.[4] Due to the coupling among electrostatic, elastic, and electrostrictive effects, local polarization fields within such nanostructures may be able to adopt unusual arrangements, including vortex-like and skyrmionic patterns.[5-7] An alternative approach to controlling the nanoscale properties of ferroelectrics involves creating ultrathin-films and superlattices (SLs) with individual component layers ranging in thickness from a few unit cells to a few nanometers.[8-10] In such heterostructures, elastic and depolarization energies differ significantly from those of bulk materials, which can dramatically change both the sequence and the nature of the exhibited ferroelectric phase transitions.[8, 11-13] For example, under certain strain conditions, some SLs have been shown to support arrays of in-plane polarization vortices.[14] Further advances in the understanding and control of ferroelectric polarization may allow other exotic polarization configurations to be experimentally realized and controlled.

Here we report a promising direction in the nanoscale control of ferroelectricity in which the formation of nanostructures is combined with heteroepitaxial SL starting materials, merging two previously independent approaches. We demonstrate that, unlike non-patterned SLs, nanostructures formed in ferroelectric/dielectric SLs possess new channels for relaxing free energy and thereby adjusting their ferroelectric properties, including the in-plane orientation of polar domain walls. This observation suggests that nanostructures in ferroelectric/dielectric SLs can be used to manipulate ferroelectric nanodomain patterns in a more general way than the specific elongated structures considered here. Such functionalities can enable a variety of applications, including data storage, optical devices, and reconfigurable electronics.[15-17] Furthermore, by elucidating the underpinnings of the SL nanostructure behavior, we gain insight into fundamental effects that can be exploited to manipulate polarization patterns at the nanoscale.

The ferroelectric SLs that serve as a starting point for the nanostructure fabrication have domain configurations that differ significantly from those observed in ultrathin ferroelectric thin films.[14, 18-19] Specifically, nanodomains formed within ferroelectric SLs have an unusually large geometric aspect ratio, with their heights (~100 nm) being many times larger than their widths (~few nm).[20-21] The peculiar geometries of the SL stripe nanodomain patterns are determined by the balance of competing domain wall and depolarization energies,[9, 19-20, 22-26] which in this particular case is heavily influenced by the presence of a large number of repeating ferroelectric/dielectric units throughout the total thickness of the SL. Their extension over a large volume makes SL nanodomain patterns far less sensitive to surface effects, including charges and dipole moments at atomic steps, which have a significant impact on the geometry of domain patterns encountered in films with few-nm thickness.[9, 26-27] This makes control of the stripe domain pattern significantly harder to achieve. Here we show that SL nanostructures provide effective alternative means for controlling the orientation of the ferroelectric nanodomains.

Discarding the thin-film SL geometry in favor of nanopatterned heterostructures is a particularly promising approach for influencing polarization at nanoscale because the shape, size, and crystallographic orientation of the nanostructure become new parameters for fine-tuning the morphology and functional behavior of its polar nanodomains. All of these parameters can drastically alter the competition among the different contributions to the total free energy of the system and thus have a significant impact on the resulting nanodomain configuration. Understanding the influence of each parameter, as well as the compounding effect of applied electric and mechanical boundary conditions on the domain pattern evolution in ferroelectric/dielectric SL nanostructures is a key step in the development of new mechanisms for controlling the location and structure of domain walls.

We focus here on the PbTiO$_3$/SrTiO$_3$ (PTO/STO) system, for which we have found that patterning a rectangular nanostructure produces specific orientations of the nanodomain patterns. In particular, the 180° stripe domain-wall configurations observed in unpatterned PTO/STO SLs still persist in the nanostructures, but have the domain pattern aligned to be parallel to the edges of the nanostructure. The synchrotron x-ray nanobeam diffraction study of the SL nanostructures reveals the presence of alignment, which leads to anisotropy in the intensity distribution of the x-ray diffuse scattering produced by the domains. A complementary computational investigation of the energetics of the nanostructure polarization patterns, using phenomenological Landau-Ginzburg-Devonshire (LGD) theory, indicates that domain wall orientation parallel to the edges of the structure yields elastic distortions that allow for the largest magnitude of the remnant polarization within the domains and in turn minimize the total energy of the system. The x-ray study also finds that the SL nanostructure is mechanically slightly distorted by the fabrication process. By including a range of applied stresses corresponding to this distortion in the computational model we show that the domain wall alignment with the geometrical features of the nanostructure is robust for the experimental widths of nanostructures considered in this study.

## Nanostructure Fabrication

The starting point for nanostructure fabrication was a thin film SL consisting of alternating layers of PTO and STO deposited using off-axis radio-frequency sputtering. The 100-nm-thick SL thin film consisted of 7 unit cells of PTO and 3 unit cells of STO, repeated 25 times, deposited on a SrRuO$_3$ (SRO) thin film on an (001)-oriented STO substrate. The deposition, structural characterization, and equilibrium domain pattern in these materials have been previously described.[21, 23, 28]

Elongated nanostructures were isolated from the surrounding area of SL heterostructure by removing areas of the SL thin film using focused-ion-beam (FIB) lithography, as illustrated in Fig. 1a. The SL was protected from ion-induced damage by first depositing a two-layer protective cap on top of the regions of the SL film in which the patterns were formed. The bottom layer of the protective cap consisted of a 130-nm-

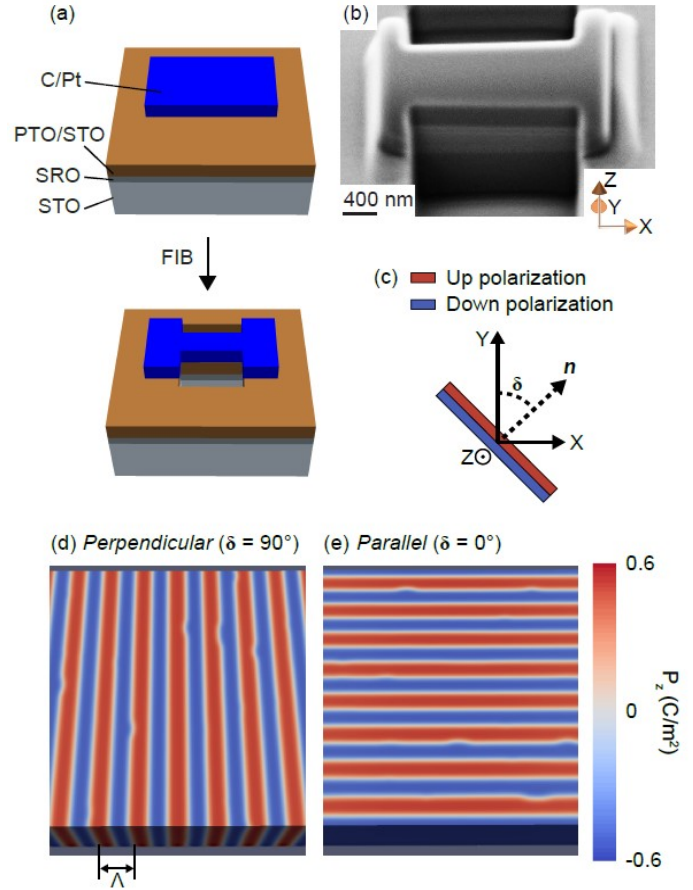

Fig. 1 (a) Schematic of nanostructure fabrication. (b) SEM image of an 800 nm-wide PTO/STO SL nanostructure created using FIB lithography. The protective cap appears as a raised region covering ridge-shaped nanostructure. The cap also extends slightly into the region above the unpatterned area of the SL at each edge of the nanostructure. (c) Definitions of parameters used to describe the domain wall orientation: the local domain-wall in-plane normal (n) and azimuthal angle (δ) with respect to the y axis. Renderings of the polarization within computational models of SL nanostructures in (d) *perpendicular* (δ = 90°) and (e) *parallel* (δ = 0°) domain configurations. The domain period (Λ) is the real-space repeat length of the domain pattern.

thick Pt layer deposited by electron-beam induced deposition. The second layer of the cap consisted of a 230-nm-thick C layer produced by Ga-ion-induced deposition. The area covered by the protective cap had dimensions of approximately 3 μm × 2 μm, extending beyond the area that was eventually milled by the focused ion beam. Regions of the cap, the SL, and underlying substrate were removed to a depth of 3 μm, isolating a ridge-shaped nanostructure. The milling was conducted using a Ga-ion accelerating potential of 30 kV and current of 50 pA in a large number of passes of the beam over each area to be removed. Synchrotron x-ray nanobeam studies of submicron-thickness STO sheets processed with a similar milling procedure have indicated that the FIB processing induces bending of the STO crystals, but does not lead to the introduction of extended defects such as dislocations.[29] Fig. 1b shows a scanning electron microscopy (SEM) image of a SL nanostructure fabricated in this way with width W=800 nm and length L=2 µm. The analysis employs a Cartesian coordinate system in which the long direction of the nanostructures is parallel to the *x* axis.

In order to describe the orientation of domains with respect to the SL nanostructure, we define a position-dependent unit vector n that is locally normal to the planes of the domain walls at all positions. The in-plane orientation ***n*** at each position is described by an azimuthal angle $\delta$ with respect to the *y* axis, as illustrated in Fig. 1c. Below we systematically compare two distinct stripe domain configurations with different domain-wall orientations, for which the polarization configurations computed using the model described below are shown in Fig. 1d and Fig. 1e. The two configurations are termed *perpendicular* and *parallel*, defined by the orientation of the domain walls relative to the edges of the nanostructures. The *perpendicular* and *parallel* configurations have $\delta$=90° and $\delta$=0°, respectively.

The SL nanostructures were probed using x-ray nanodiffraction at station 26-ID-C of the Advanced Photon Source of Argonne National Laboratory, using the arrangement illustrated in Fig. 2a. The x-ray beam employed in this study had a photon energy of 9 keV, at which the x-ray absorption length in the SL material is far greater than the SL thickness. The x-ray beam thus penetrates the entire thickness of the SL, and provides a sensitive probe of the nanoscale domain orientation and structural distortion.[30] The x-rays were focused to a 50 nm full-width-at-half-maximum spot using a 160 µm-diameter Fresnel zone plate. A 60 µm-diameter center stop and an order sorting aperture were used to attenuate the unfocused beam and x-rays focused to higher orders. The convergence of the focused x-ray nanobeam was 0.26° at the focal spot. X-ray diffraction patterns were collected using a charge coupled device detector consisting of a 1024×1024 array of 13 µm square pixels. The focused x-ray beam propagated in a plane illustrated in Fig. 2a, defined by the length of the SL nanostructure and its surface normal, the *x-z* plane. The nanostructures were located by creating spatial maps of the Pb M-edge x-ray fluorescence of the PTO/STO SL to identify the patterned regions in which the SL had been removed. Experiments were conducted under diffraction conditions near the (002) PTO/STO SL Bragg reflection near the Bragg angle $\vartheta_B$=20.11°. The corresponding out-of-plane wavevector $q_z$ is 3.14 Å$^{-1}$.

The key structural feature of the SL diffraction pattern is the 0th order Bragg reflection of the SL. This reflection appears at wavevector corresponding to the average periodicity of the SL.[31] In addition, a ring of diffuse x-ray scattering intensity arises from the ferroelectric striped nanodomains. The reciprocal-space radius of the domain diffuse scattering ring acquired in the unpatterned PTO/STO SL region was 0.097 Å$^{-1}$. This radius corresponds to a stripe nanodomain period $\Lambda$ of 6.5 nm. The distribution of intensity around ring of domain diffuse scattering provides insight into the orientation of domain walls within the striped domain pattern. The epitaxial synthesis of an unpatterned film does not lead to a preferred orientation of the domain walls.[9-10, 21, 26] As a result, the diffuse scattering intensity from the nanodomains in the unpatterned film is distributed approximately uniform ring intensity of domain diffuse scattering in the $q_x$-$q_y$ plane, schematically shown as a toroid in Fig. 2b.

## Results and Discussion

In cases in which there is a non-uniform distribution of the orientations of the domain walls there is also a non-uniform distribution of scattered x-ray intensity around the ring of domain diffuse scattering. Crucially, the intensity at each azimuthal angle around the ring of diffuse scattering in reciprocal space arises from domains with matching directions of the in-plane real-space vector normal to the domain wall.

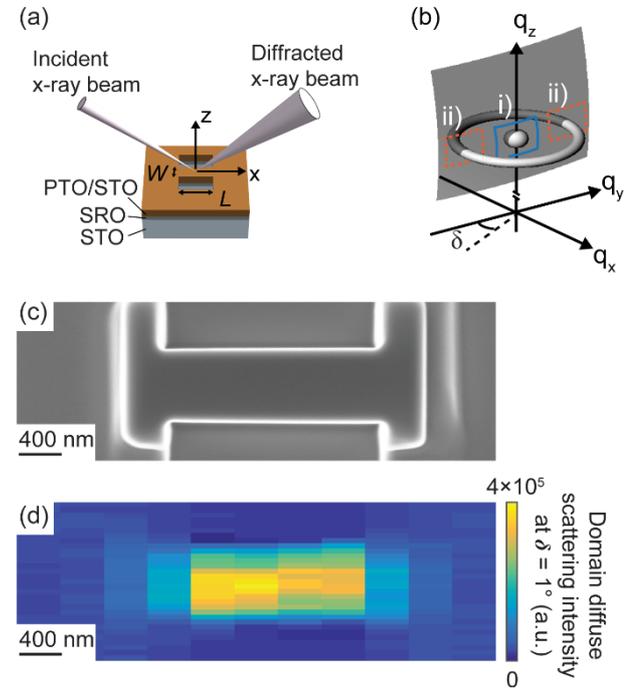

Fig. 2 (a) X-ray nanodiffraction geometry including the PTO/STO SL thin film heterostructure, underlying SRO layer, and STO substrate. (b) Geometry of reciprocal space near the (002) x-ray reflection of the PTO/STO SL. When the diffraction experiment matches the Bragg condition for the SL, at $\Delta\vartheta$ = 0, the Ewald sphere intersects i) Bragg reflection and ii) the ring of domain diffuse scattering. At other values of $\Delta\vartheta$ the Ewald sphere intersects the ring of domain diffuse scattering at a different value of $\delta$. (c) SEM image of an 800 nm-wide SL nanostructure. (d) Map of domain diffuse scattering intensity acquired at $\Delta\vartheta$ = 0.03°, corresponding to $\delta$ = 1° in the area shown in (c). The intensity within the nanostructure is higher than in the unpatterned region by a factor of 7.

Based on the experimental arrangement shown in Fig. 2a and Fig. 2b, the domain diffuse scattering intensity at small values of $\delta$ can be expected to be enhanced when the domain walls are parallel to the mechanically milled edges of the SL nanostructure. An anisotropy of the ring of magnetic domain diffuse scattering has similarly been observed in magnetic striped domain systems.[32]

The x-ray scattering Ewald sphere can be adjusted to intersect the domain diffuse scattering at different values of the domain orientation angle $\delta$ by varying the incident angle of the focused x-ray beam. The value of $\delta$ depends on the angular difference $\Delta\vartheta$ between the incident angle of the x-ray beam and the Bragg angle $\vartheta_B$ of the SL structural reflection:†

$$\delta(\Delta\vartheta) = \sin^{-1}\left[\frac{\Lambda}{\lambda}\left\{\sqrt{1-\left(2\sin\vartheta_B - \sin(\vartheta_B+\Delta\vartheta)\right)^2} - \cos(\vartheta_B+\Delta\vartheta)\right\}\right]$$

A derivation of this relationship is given in the supplementary materials.† The maximum value of $\Delta\vartheta$ at which the Ewald sphere can geometrically intersect the ring of domain diffuse scattering is 1.77°. For magnitudes of $\Delta\vartheta$ less than approximately 1°, as in this study, $\delta(\Delta\vartheta) \approx 2\frac{\Lambda}{\lambda}\sin(\vartheta_B)\Delta\vartheta$, or $\delta(\Delta\vartheta) = 32.4\,\Delta\vartheta$.

Two key observations arise from the x-ray nanobeam characterization of the nanostructures. First, the domain diffuse scattering shows that the formation of the nanostructure induces anisotropy in the distribution of domain diffuse scattering. A scanning electron microscopy image of an SL nanostructure with a width of 800 nm is shown in Fig. 2c. The small-$\delta$ domain diffuse scattering intensity from the nanostructure was collected with $\Delta\vartheta=0.03°$, corresponding to $\delta=1°$. The domain diffuse scattering intensity as a function of position within and near the SL nanostructure is shown in Fig. 2d. As is apparent from the map in Fig. 2d, the nanostructure produced a significantly enhanced domain diffuse scattering intensity at small values of $\delta$, a factor of 7 higher than the domain diffuse scattering in the unpatterned region. The local enhancement of the diffuse scattering provides an initial indication that domain walls in the SL nanostructure are aligned with the edges of the structure. The map presented in Fig. 2d also indicates that there are regions of comparatively lower domain diffuse scattering intensity near the edges of the nanostructure. This apparent reduction in intensity arises from an artifact in the x-ray imaging conditions due to a combination of a deviation from the x-ray focal condition at this location and an azimuthal misalignment of the x-ray nanobeam footprint with the long axis of the nanostructure.

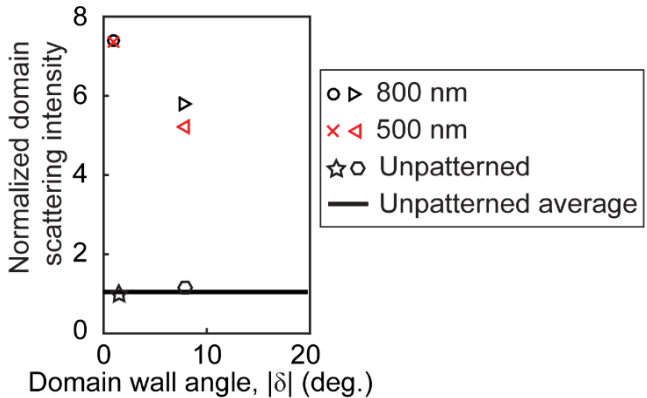

Fig. 3 Normalized domain diffuse scattering intensities as a function of azimuthal angle $\delta$. Domain scattering intensities were measured in unpatterned regions and in 500 nm- and 800 nm-wide SL nanostructures at domain normal angles $\delta=1°$ and 8°.

Further information about the azimuthal distribution of domain diffuse scattering intensity was obtained by comparing the domain diffuse scattering intensities measured at $\Delta\vartheta=0.03°$ and 0.25°, corresponding to $\delta=1°$ and 8°. The azimuthal dependence of the domain diffuse scattering intensities for SL nanostructures with widths of 500 nm and 800 nm are plotted in Fig. 3, along with the intensities acquired at the same angles in unpatterned regions of the SL appearing at the left and right edge of Fig. 2c. The intensities in Fig. 3 are plotted on a scale normalized by the average intensity of the measurements at the two orientations in the unpatterned region. On this scale, the normalized intensity of the unpatterned regions is close to 1. The domain diffuse scattering intensities in SL nanostructures have a high value at low $\delta$ and a slightly lower value at the larger $\delta$. The domain walls are thus preferentially aligned along the mechanical edges of the nanostructures. Under the assumption that the distribution of domain wall directions is symmetric around $\delta=0$, a fit of a normal distribution of domain wall orientations gives a FWHM of 20°. This angular width indicates that the domain walls are parallel to the long edge of the nanostructure with deviations in their orientations of approximately ±10°.

The presence of the capping layer makes domain wall imaging by piezoelectric force microscopy microscopy (PFM) impossible because the electrical contact between a probe tip and the ferroelectric is interrupted. Similarly, the formation of the thin sections of the SL that would be required for transmission electron microscopy could potentially perturb the original conditions leading to domain alignment. We have thus not attempted to use either of these techniques to probe the obtained domain wall configurations.

We have quantitatively considered, and ultimately discarded, three artifacts that could in principle lead to the enhanced domain diffuse scattering intensity observed in the SL nanostructures without domain alignment. First, an increase in the total number of domain walls within the region illuminated by the x-ray beam, corresponding to a decrease in domain period, could lead to increased domain diffuse scattering. A comparison of the domain diffuse scattering from patterned and unpatterned regions shows that the domain period in SL nanostructures is 3% smaller than the unpatterned region. This would produce 6% increase in intensity within the nanostructures because the x-ray diffuse intensity is proportional to the square of the total number of domain walls.[19, 25] The expected intensity change due to the domain period is thus much smaller than the measured intensity difference between patterned and unpatterned areas.

A second possible artifact is related to the possibility that SL nanostructures could have reduced domain wall width and thus higher x-ray scattering in comparison with the unpatterned region. This possibility can be evaluated using the domain diffuse scattering intensities derived as a function of domain-wall width.[33] Assuming that the domain wall width is zero in the SL nanostructure, the observed intensity enhancement occurs only when the domain wall width reaches 16 nm in the unpatterned region, which is an unphysically large value.[34, 35]

A final artifact would arise if the areas occupied by up and down polarizations within the nanodomain period were different in the SL nanostructure than in the unpatterned region. The intensity of the diffuse scattering depends on the up-down domain fraction and reaches a maximum with equal populations of up and down polarizations.[19, 25] An up-polarization fraction $p_{up}$=0.5 in the SL nanostructure and $p_{up}$=0.1 in the unpatterned region would lead to the observed intensity enhancement. The change in up-down polarization fraction, however, is unreasonable because we expect $p_{up}$=0.5 in the unpatterned region, as been observed in previous studies of PTO/STO SLs grown on STO substrates.[22] Careful consideration of the experimental artifacts thus shows that none of these contributions accounts for the observed intensity enhancement.

The x-ray nanobeam diffraction study also reveals that the SL nanostructures are elastically distorted as a result of the FIB patterning process. The distortion leads to a position-dependent angular shift of the SL Bragg reflection due to the local tilting of the lattice planes. A map of the tilt of the crystal lattice within the SL nanostructures, shown in Fig. 4a, indicates that the lattice is tilted by up to 0.07° with respect to the average orientation. As expected from the two-dimensional symmetry of the epitaxial growth of the SL heterostructure, the areas of the scan in Fig. 4b away from the nanostructure reveal that there is no systematic variation in the tilt in the unpatterned regions.

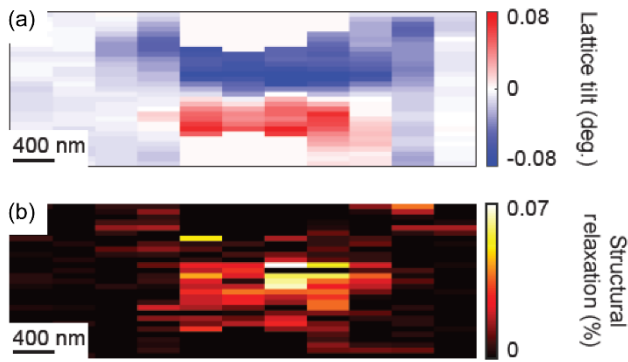

Fig. 4 (a) Lattice tilt within the 800 nm-wide SL nanostructure shown in Fig. 2c. (b) In-plane structural expansion computed under the assumption that the strain arises only from bending of the nanoscale sheet.

The distortion of the lattice of the PTO/STO nanostructure is an important parameter because of the coupling between stress and polarization in ferroelectrics. The x-ray nanodiffraction data provides insight into the magnitude of the distortion of the lattice in two ways. First, the curvature of the nanostructure apparent in Fig. 4a introduces a change in the average lattice parameter. If we assume that the strain arises only from the bending of the SL, the magnitude of the in-plane strain can be approximated as an expansion with magnitude on the order of $t/2R$, where $t$ is the film thickness and $R$ is the radius of curvature. The average radius of curvature derived from Fig. 4a is $R$=0.17 mm, which gives an average in-plane expansion of 0.03%. The local curvature calculated by computing the numerical derivative of the lattice tilt was used to calculate the local values of the relaxation, as shown in Fig. 4b. A second estimate of the distortion of the PTO/STO nanostructure was obtained by analyzing the shift in the magnitude of PTO/STO 002 Bragg reflection wavevector between a 500-nm-wide SL nanostructure and an unpatterned region (see Supplemental Material).[36] The strain along the surface normal direction obtained using this method was $\varepsilon_z$=-0.08%. The strain in the plane of the surface can be determined from the surface normal strain using a relationship developed for bent single-crystalline sheets, $\varepsilon_x$=(ν-1)/ν $\varepsilon_z$.[37, 38] We use ν=0.3125, the value of the Poisson ratio for PTO because the mechanical properties of the PTO/STO SL are not yet known.[39, 40] The measured strain along the surface normal thus corresponds to a strain in the $x$ direction of $\varepsilon_x$=0.18%, similar to the values derived from the curvature. The magnitude of the strain is also similar to what we have previously observed in STO sheets formed using FIB.[29] We note that the strain observed here is slightly less than has been recently reported in Si or Au nanocrystals exposed to the FIB because the top of the SL is protected with a capping layer and is not directly milled.[41-43] The observed in-plane relaxation provides an important parameter for demonstrating the validity of the computational results described below.

## Computational Mesoscopic Model

A range of computational methods, including density functional theory and phase field modeling, can be used to predict the polarization configuration of nanoscale ferroelectrics.[22, 44-46] We have conducted a computational study using a phase-field method to evaluate the energetics of different domain wall arrangements in the SL nanostructures. Previous studies have shown that PTO/STO SLs in which the PTO component dominates the SL repeating unit adopt a low-energy configuration with a near-homogenous polarization of intermediate magnitude[28, 47] and that the STO layers are strongly polarized.[8, 48] Since the polarization is almost constant throughout the PTO/STO repeating unit of the SL, to simplify the simulations we adopted a model configuration in which the SL was represented by a uniform ferroelectric material.

Three computational models of different sizes were constructed, with SL nanostructure dimensions of 40×40×25, 60×60×25, and 80×80×25 nm³, respectively. Periodic boundary conditions were applied along only the $x$ direction (with the $y$ and $z$ directions kept finite), effectively making the system infinitely long along the length of the nanostructure, while remaining finite in the $yz$ plane. The models are considerably smaller than the dimensions of the experimentally fabricated nanostructures along the patterned $y$ and $z$ directions. As we show below, the results produced by these computational models of smaller nanostructures can be extrapolated to predict the behavior of nanostructures with the experimental widths of 500 and 800 nm.

The behavior of coupled polarization, electrostatic potential and elastic stress fields in the SL nanostructure was investigated with the open-source package Ferret[7, 49] based on the Multi-physics Object-Oriented Simulation Environment (MOOSE) framework.[50] The temporal evolution of the polarization field vector ***P*** within the system at each location was described by the time-dependent Landau-Ginzburg-Devonshire (TDLGD) equation:

In the TDLGD simulation, the system evolves through changes that reduce free energy until a stable configuration representing a global or local minimum is reached. The simulations were terminated when the change in total energy was less than 0.1% per time step. Two domain configurations

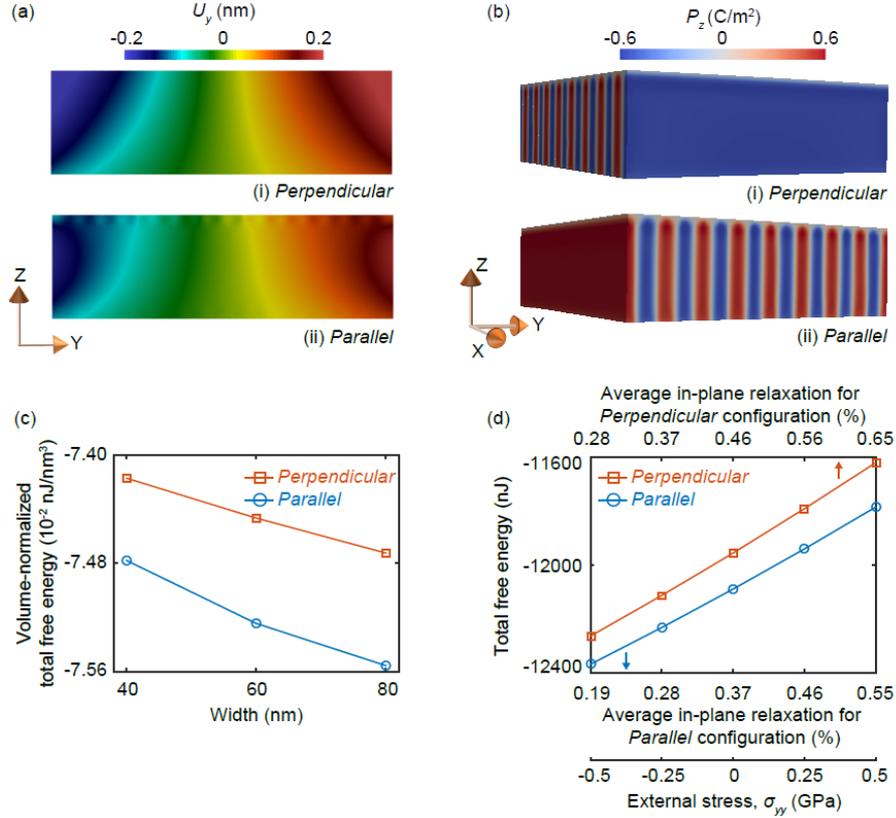

Fig. 5 Maps of (a) the *y*-component of the displacement field ($U_y$) (b) *z*-component of the polarization ($P_z$) in the (i) *perpendicular* and (ii) *parallel* configurations in an 80 nm-wide SL nanostructure. (c) Nanostructure-width-dependent volume-normalized total free energy. Extrapolation to larger widths indicates that the *parallel* configuration remains energetically favorable at the experimental nanostructure widths. (d) Free energy as a function of external stress applied along the *y* direction $\sigma_{yy}$, indicating that the *parallel* configuration is favorable under experimentally applied stresses arising from the lithography processes.

$$-\gamma \frac{\partial \mathbf{P}}{\partial t} = \frac{\delta}{\delta \mathbf{P}} \int_V f(\mathbf{P}) d^3 r$$

Here, $f(\mathbf{P})$ is the local LGD free energy density comprised of ferroelectric bulk, electrostatic, domain wall, elastic, and coupled polarization-strain (electrostrictive) energy terms, and $\gamma$ is a time constant that describes the mobility of domain walls.[51] We arbitrarily set $\gamma=1$ because we were not interested in specifics of evolution trajectories of the system, but rather only in the final steady-state system configurations and energies. Periodicity was enforced on all the system variables (polarization, elastic displacements, and electrostatic potential fields) along the *x* axis. The nanostructure was surrounded by vacuum regions along the non-periodic directions. At each time step, the TDLGD, elastic stress-divergence and electrostatic Poisson equations were solved under assumption that elastic and electrostatic fields relax at a much faster rate than polar distortions.[52] Further details of the simulation, including materials parameters, are provided in the Supplemental Materials.† were considered as initial conditions of the simulation. The polarization field within the PTO film was pre-biased to form 180° stripe domain patterns with domain walls being either *perpendicular* or *parallel* to the milled edges of the nanostructure. An initial polarization modulation with domain period Λ matching the experimental values for the considered PTO/STO SLs was also imposed as part of the initial conditions of the computation. The material parameterization used in the computation resulted in a simulated magnitude of the polarization of 0.6 C/m² in each up and down stripe domain. For both *perpendicular* and *parallel* initial conditions, the ensuing domain configurations retained their original orientations with respect to the nanostructure surfaces and evolved to distinct local energy minima. No conversion of *perpendicular* configuration into parallel configuration or vice-versa was observed.

The simulations revealed that, for both *perpendicular* and *parallel* configurations, the nanostructure elastically relaxes along the non-periodic directions *y* and *z*. The *y*-component of the displacement field ($U_y$) for the relaxed state reached in *perpendicular* and *parallel* configurations in an 80-nm-wide nanostructure is shown in Fig. 5a. The displacement field map in Fig. 5a shows that both *perpendicular* and *parallel* configurations relax outward toward the edges of the structure. These expansions are slightly different for the two configurations. The displacement for the *perpendicular* configuration is 0.20 nm at each edge and the displacement for the *parallel* configuration is slightly smaller, 0.18 nm. The *parallel* configuration also develops a corrugated surface with the same periodicity as the domain pattern. The corrugated surface originates from the in-plane compression of the nanostructure at domain walls in which the in-plane polar vectors are under compressive in-plane strain.[22] The relaxed states of both *perpendicular* and *parallel* configurations retain the periodic striped polarization pattern as shown in Fig. 5b.

The simulations also revealed that the *perpendicular* and *parallel* configurations have large differences in the bulk and coupled terms of the free energy. Both the bulk and coupled polarization-strain energy terms are consistently considerably lower in the *parallel* configuration, due to the greater magnitude of the z-component of the polarization obtained in that configuration. The larger z-component of the polarization results from a compression at the near-surface domain walls. A similar near-surface relaxation at domain boundaries has been observed in simulations of ultrathin PTO layers.[53] The lower bulk and polarization-strain energy counterbalances the higher elastic energy of the *parallel* configuration. As a result of the difference between the bulk and coupled contributions between the two configurations, the total energy of the *parallel* configuration is lower than that of the *perpendicular* configuration. Based on the insights from Fig. 5, we thus can summarize the mechanism of the alignment. The *parallel* configuration permits the in-plane compression at the near-surface domain walls which gives rise to the more favorable free energy of the *parallel* nandomain pattern.

Other components of the system free energy have a smaller influence than the difference in bulk and coupled contributions in the *perpendicular* and *parallel* configurations. For both *perpendicular* and *parallel* configurations, the electrostatic energy terms are negligibly small in comparison with other energy contributions. Both domain configurations thus provide adequate compensation of the depolarization fields stemming from the surface charges. Furthermore, both configurations have the same area of domain walls, and have similar domain periods, leading to only minor differences in the domain-wall formation energies, with the energy of the *perpendicular* configuration being slightly lower.

The results of the calculations can be extrapolated to provide insight into the behavior of the experimentally probed SL nanostructures. Fig. 5c shows the dependence of the total energy of the *perpendicular* and *parallel* configurations on the width of the nanostructures. The total energy of the *parallel* configuration is always lower than that of the *perpendicular* configuration in all of the model sizes considered. In the limit of very large nanostructure width, physical intuition suggests that both configurations should become identical and thus have the same energy. The extrapolation of the data in Fig. 5c suggests that this convergence will occur for lateral sizes that are much larger than those of the experimental samples of 500 nm and 800 nm. Therefore, we conclude that, in the experimental nanostructures, the *parallel* configuration is energetically more favorable than *perpendicular* configuration, and the observed differences between the energy terms in these two configurations are experimentally meaningful at these sizes.

In addition to the factors associated with the initial domain geometry, it is also important to consider the potential role of the structural distortions that arise from the fabrication of the SL nanostructures. The x-ray measurements in Fig. 4 show that the SL nanostructures are externally stressed by effects linked to FIB. In order to examine whether the energetics of the domain-wall configurations could be influenced by external stresses, we have repeated the simulation under a range of applied mechanical boundary conditions. The energies of the *perpendicular* and *parallel* configurations subjected to applied external stresses ($\sigma_{yy}$) up to ±0.5 GPa along the y axis were obtained using the computational method described above. The stress dependence of the computed free energy is shown in Fig. 5d. The applied stress produced a spatially varying distortion for which there is no unique value of the in-plane strain. Fig. 5d shows the average in-plane relaxations, which range from 0.28% to 0.65% or from 0.19% to 0.55%, for *perpendicular* and *parallel* configurations, respectively. The simulations showed that the *parallel* configuration remains the energetically preferable configuration throughout the whole range of probed external stresses. The range of relaxations considered in the simulations spans (and far exceeds) the experimentally observed average in-plane strain of less than 0.1%. Therefore, we conclude that elastic artifacts associated with the lithographic processes do not substantially alter the domain-wall configuration.

## Conclusions

The development of ferroelectric/dielectric SL nanostructures represents a new direction in nanostructured ferroelectrics that provides strategies for controlling ferroelectricity. From a fundamental perspective, structural and computational studies of SL nanostructures reveal effects that can be used to manipulate nanodomain patterns. In particular, the coupling of mechanical and ferroelectric properties results in the alignment of polarization nanodomains in PTO/STO SL nanostructure. The LGD calculations show that this alignment mechanism is a mesoscale effect that emerges in ferroelectric nanostructures incorporating internal SLs. A *parallel* domain configuration is energetically more favorable than the *perpendicular* configuration due to the greater magnitude of the polarization in the *parallel* configuration. This larger polarization in turn emerges because of the in-plane structural distortion present in the nanostructure. Crucially, this alignment does not depend on stresses introduced by the lithography process.

More generally, the combination of nanostructure formation and SL heterostructuring allows the manipulation of domain walls in SLs and similar materials through the control of mechanical boundary conditions. This approach has the potential to produce nanostructures with controlled locations of exotic local polarization patterns such as vortex-like configurations,[22, 54] flux-closure domains,[55] or conductive domain boundaries.[56] Manipulating the orientation and position of these complex polarization states via nanostructuring provides a new dimension for understanding and eventually applying functional properties of nanoscale ferroelectrics.

## Conflicts of interest

There are no conflicts to declare

## Acknowledgements

JP, QZ, AP, and PGE acknowledge support from the U.S. DOE, Basic Energy Sciences, Materials Sciences and Engineering, under contract no. DE-FG02-04ER46147 for the x-ray scattering studies and analysis. Use of the Center for Nanoscale Materials and the Advanced Photon Source, both Office of Science user facilities, was supported by the U.S. Department of Energy, Office of Science, Office of Basic Energy Sciences, under contract no. DE-AC02-06CH11357. JM acknowledges funding support from the U.S. Department of Energy, Office of Science, Office of Workforce Development for Teachers and Scientists, Office of Science Graduate Student Research (SCGSR) program. The SCGSR program is administered by the Oak Ridge Institute for Science and Education (ORISE) for the DOE. ORISE is managed by ORAU under contract number DE-SC0014664. OGH acknowledges support from the US DOE Office of Science, Basic Energy Sciences Division of Materials Science and Engineering. JM, OGH, and SMN also acknowledge the computing resource support provided on Blues, a high-performance computing cluster operated by the Laboratory Computing Resource Center at Argonne National Laboratory, and on the Hornet cluster, hosted by the Taylor L. Booth Engineering Center for Advanced Technology, located at the University of Connecticut at Storrs. The authors acknowledge use of facilities and instrumentation supported by NSF through the University of Wisconsin Materials Research Science and Engineering Center (DMR-1121288 and DMR-1720415).

## Notes and references